\documentclass[twocolumn,secnumarabic,amssymb,nobibnotes,aps,pra,10pt,superscriptaddress]{revtex4-1}

\usepackage[T1]{fontenc}
\usepackage[utf8]{inputenc}
\newcommand{\latin}[1]{{\emph{#1}}}

\usepackage{graphicx}
\usepackage{color}
\usepackage{physics}

\usepackage{setspace}

\usepackage{stmaryrd}
\usepackage{bbm} 
\newcommand{\N}{\mathbb{N}} 
\newcommand{\diff}{\mathop{}\mathopen{}\mathrm{d}} 
\newcommand{\identite}{\ensuremath{\mathbbm{1}}}

\usepackage[colorlinks]{hyperref}

\hypersetup{%
pdftitle={Quantum description of timing-jitter for single photon ON/OFF detectors},
pdfauthor={Gouzien, Élie and Fedrici, Bruno and Zavatta, Alessandro and Tanzilli, Sébastien and D'Auria, Virginia},
pdfsubject={We propose here a theoretical model based on POVM density formalism able to explicitly quantify the effect of timing-jitter for a typical class of single photon detector. We apply our model to some common experimental situations.},
pdfkeywords={single photon detector, timing-jitter, quantum, POVM}%
}

\begin{document}

\title{Quantum description of timing-jitter for single photon ON/OFF detectors}

\author{Élie Gouzien}
\affiliation{Université Côte d'Azur, Institut de Physique de Nice (INPHYNI), CNRS UMR 7010, Parc Valrose, 06108 Nice Cedex 2, France}
\author{Bruno Fedrici}
\affiliation{Université Côte d'Azur, Institut de Physique de Nice (INPHYNI), CNRS UMR 7010, Parc Valrose, 06108 Nice Cedex 2, France}
\author{Alessandro Zavatta}
\affiliation{Istituto Nazionale di Ottica (INO-CNR) Largo Enrico Fermi 6, 50125 Firenze, Italy}
\affiliation{LENS and Department of Physics, Universitá di Firenze, 50019 Sesto Fiorentino, Firenze, Italy}
\author{Sébastien Tanzilli}
\affiliation{Université Côte d'Azur, Institut de Physique de Nice (INPHYNI), CNRS UMR 7010, Parc Valrose, 06108 Nice Cedex 2, France}
\author{Virginia D'Auria}
\email{virginia.dauria@inphyni.cnrs.fr}
\affiliation{Université Côte d'Azur, Institut de Physique de Nice (INPHYNI), CNRS UMR 7010, Parc Valrose, 06108 Nice Cedex 2, France}

\begin{abstract}
In the context of ultra-fast quantum communication and random number generation, detection timing-jitters represent a strong limitation as they can introduce major time-tagging errors and affect the quality of time-correlated photon counting or quantum state engineering.
Despite their importance in emerging photonic quantum technologies, no detector model including such effects has been developed so far.
We propose here an operational theoretical model based on POVM density formalism able to explicitly quantify the effect of timing-jitter for a typical class of single photon detector.
We apply our model to some common experimental situations.
\end{abstract}
\maketitle

\section{Introduction}
Quantum communication stands as one of the most promising applications of quantum optics with an increasing number of encouraging out-of-the-laboratory implementations~\cite{JWPanRelayNature, Satellite}.
The quest for competitive quantum photonic systems, compatible with existing standard technology, has promoted huge developments concerning both photonics sources~\cite{duccireview, Tanzillireview} and detectors~\cite{Poliakov2011, HadfieldNatPhot2009, Marsili2013-8inCalandri}.
Nevertheless, a critical point still lies in experiments' operation rates.
Time multiplexing technics allow in principle to pump photonic sources at rates on the order of GHz~\cite{Yamamoto_2007_10GHz, Sasaki_2014_SciRep, Lutfi2015, JWPanSwapping}.
However, in practical realisations, a strong restraint to ultra-fast regime comes to timing limitations at the detection stage.
Dead-times after each detection event restrict the maximum rate at which output signals can be delivered~\cite{HadfieldNatPhot2009}.
At the same time, timing-jitters limit the experimental temporal resolution by introducing random fluctuations on the times $T$ at which output signals are delivered, even in the particular case of a single photon perfectly temporally localized (see Fig.\,\ref{fig:definition}).
If not circumvented, this effect can be extremely detrimental in high speed regimes, as it can cause counts relative to different experiment clock cycles to become temporarily indistinguishable~\cite{HadfieldNatPhot2009, IEEE2010}.

Fast and accurate time-tagging is mandatory in multiple operations, such as quantum teleportation~\cite{JWPanRelayNature}, quantum state engineering~\cite{Lutfi2015} and quantum random number generation~\cite{AmayaPRA2015}.
In anticipation to further technological advances, as well as in the perspective of promoting novel quantum communication protocols, it is thus of the utmost importance to correctly describe the effects of detectors' timing performances.
Despite a huge number of experiments reporting measured time response of different photon-counting devices~\cite{HadfieldNatPhot2009, Calandri2016, Tosi2014-5inCalandri}, these effects have never been fully included in a quantum detection model.
In this paper, we address this point with a theoretical model providing an operational and explicit description of a standard single photon detectors affected by non negligible timing-jitter, in presence of dead-time, and not restricted to single-photon states.
Its impact goes from quantum technologies to any single-photon based application where evaluating the influence of detector timing uncertainties can allow in principle to amend their effect.

We adopt the formalism of positive operator-valued measurements (POVM)~\cite{MatteoParisBook}.
This approach has been vastly exploited to describe different types of measurement apparatus~\cite{WalsmleyContiPOVMDiagonale, Schnabel2015, vanEnkPRA2017timedependentspectrum}, together with relevant figures of merit~\cite{vanEnkJoPC2017Photodetectorfiguresmerit}, as well as to experimentally investigate the characteristics of unknown detectors~\cite{WalsmleyNatPhot, DAuriaPRL2011, Brida2012, MaOpex2016, Grandi2017}.
We will focus on ON/OFF devices, \latin{i.e.\@} single photon detectors with no photon-number resolving capabilities.
These systems represent the vast majority of available single photon-counters (including avalanche photodiodes)~\cite{Poliakov2011} and our study can be generalized to multiplexed schemes~\cite{WalsmleyContiPOVMDiagonale, ONOFFCorrelation}.
Although including it is not complicated, to simplify our treatment, we chose to disregard the effect of dark-counts: their action is negligible in current ultra-fast operations involving state-of-the-art low-noise detectors, such as superconducting devices~\cite{HadfieldNatPhot2009}, as well as in all coincidence counting experiments.

\section{Theoretical model}

\begin{figure}[htbp]
\hspace*{-0.35cm}
\centering
\includegraphics[width=.7\columnwidth]{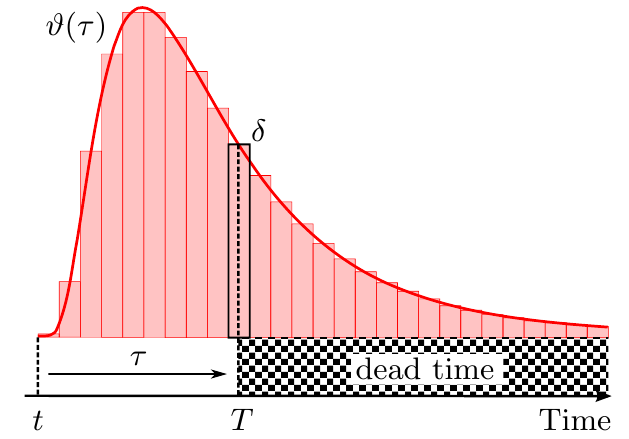}
\caption{\label{fig:definition}Typical timing parameters and response function in the ideal case of an ON/OFF detector hit at time $t$ by a perfectly temporally localized input photon and delivering an output signal at time $T$. To describe the timing-jitter effect, we introduce the random delay, $\tau=T-t$, distributed according to $\vartheta(\tau)$. The shape of $\hat{\pi}_{\text{on}}(T)$ depends on the combined effects of $\vartheta(\tau)$, dead-time, and detection efficiency, $\eta$.
}
\end{figure}

In standard treatments neglecting timing-jitter effects, given an arbitrary optical state described by the density matrix $\hat{\rho}$ impinging an ON/OFF detector, only two possible responses can be registered: ON, \latin{i.e.\@} ``at least one photon detected'', and OFF, \latin{i.e.\@} ``no photon detected''.
This situation formally corresponds to two POVM operators, $\hat{\Pi}_{\text{on}}$ and $\hat{\Pi}_{\text{off}}=\hat{1}-\hat{\Pi}_{\text{on}}$~\cite{MatteoParisBook}.
By definition of POVM, the probability of obtaining the ON result can be computed as $P_{\text{on}}=\Tr[\hat{\Pi}_{\text{on}}\hat{\rho}]$ and analogously for OFF\@.

When timing-jitter is taken into account, for an arbitrary optical input $\hat{\rho}$, a richer panel of ON events is \latin{a priori} available, corresponding each to a detection signal delivered at time $T$.
The values of $T$ are, in principle, infinite and continuously distributed.
The probability that the detector delivers an output signal at a time between $T$ and $T+\diff{T}$ ($\diff{T}$ being infinitesimal) can be generally expressed as $p_{\text{on}}(T,\hat{\rho})\diff{T}$, with $p_{\text{on}}(T,\hat{\rho})$ a density probability function.
To describe this effect, we start by considering, at the detector input, a single photon temporally localized at time $t$ and express the timing-jitter in terms of the random delay $\tau=T-t$, and associated density probability function, $\vartheta(\tau)$ (see Fig.\,\ref{fig:definition}).
Due to causality, $\vartheta(\tau<0)=0$, while its actual shape depends on the detector and is in principle inferable from experiments~\cite{HadfieldNatPhot2009, IEEE2010}.
The case of a detector with no timing-jitter is obtained in the limit of $\vartheta(\tau)$ converging to a Dirac delta distribution.

In order to deal with the continuous set of possible measurement results~\cite{POVMDensity}, we describe the detector in terms of a \emph{density} of POVM ON operators, $\hat{\pi}_{\text{on}}(T)$, such that the function $p_{\text{on}}(T,\hat{\rho})$ can be defined as:
\begin{equation}
p_{\text{on}}(T,\hat{\rho}) = \Tr[\hat{\pi}_{\text{on}}(T)\hat{\rho}].
\label{definizione}
\end{equation}
Standard ON/OFF POVMs, obtained when disregarding the temporal degrees of freedom, can be retrieved as $\hat{\Pi}_{\text{on}}=\int \hat{\pi}_{\text{on}}(T)\diff{T}$ and $\hat{\Pi}_{\text{off}}=\hat{1}-\int \hat{\pi}_{\text{on}}(T) \diff{T}$.
Moreover, in order to mimic many experimental situations, the continuous set of temporal outcomes can be conveniently discretized by introducing a partition on the values of $T$ over intervals $\delta$ based on the temporal resolution of the device electronics (Fig.\,\ref{fig:definition}).
This way, a finite number of possible outputs is obtained and, correspondingly, a finite set of POVMs, $\{\hat{\Pi}_{\text{on}}(T\in\delta), \hat{\Pi}_{\text{off}}(T\in\delta)\}$, where, $\forall\delta$, $\hat{\Pi}_{\text{on}}(T\in\delta) = \int_{\delta} \hat{\pi}_{\text{on}}(T) \diff{T}$, and $\hat{\Pi}_{\text{off}}(T\in\delta)=\hat{1}-\hat{\Pi}_{\text{on}}(T\in\delta)$.
We stress that for the definition of both ON and OFF operators, we use $\hat{\pi}_{\text{on}}(T)$.
Conceptually, this arises from the fact that as the OFF result is, by definition, given by the absence of ON events between $T$ and $T+\diff{T}$, its probability is close to 1 when considering an arbitrarily short $\diff{T}$ and, accordingly, cannot be described by a finite density function.

The explicit expression of $\hat{\pi}_{\text{on}}(T)$ can be found by considering photon-counting devices as phase-insensitive detectors~\cite{WalsmleyContiPOVMDiagonale}.
The POVMs are thus linear combinations of diagonal \emph{projectors} over basis states, $\hat{\rho}_k$, and read
\begin{equation}
\hat{\pi}_{\text{on}}(T) = \sum_{k\in \N}p_{\text{on},k}(T)\hspace{0.05cm}\hat{\rho}_k,
\label{eq:pi_diagonal}
\end{equation}
where $p_{\text{on},k}(T)$ represents the probability density function that the detector generates a signal ON at a time $T$ when hit by the state $\hat{\rho}_k$.
As the photons impinging the detector may not be perfectly simultaneous, we exploit a temporal multimode formalism~\cite{SilberhornMultimode}
and take basis states $\hat{\rho}_k$ containing $k$ temporally localized photons, each hitting the detector at a time $t_j, j \in \left\llbracket 1, k \right\rrbracket$:
\begin{equation}
\hat{\rho}_k=\ketbra{1_{t_1}}\otimes \ketbra{1_{t_2}}\otimes \dots \otimes \ketbra{1_{t_k}},
\label{rhok}
\end{equation}
where the $\ket{1_{t_j}}$ are defined from the creation operators $\hat{a}^\dagger(t_j) \ket{0}$.
Different $t_j$ correspond to different temporal modes, accordingly $[ \hat{a}\left( t_i \right),\hat{a}\left( t_{j} \right)] = 0$ and $[ \hat{a}\left( t_i \right), a^\dagger\left( t_{j} \right)] = \delta\left( t_i - t_j \right)$~\cite{SilberhornMultimode}.
In the special case of $k$ perfectly simultaneous photons ($t_j=t$, for all $j$), $\hat{\rho}_k$ reduces, up to a normalization factor, to $\ketbra{k}$, \latin{i.e.\@} to projectors over a single temporal mode Fock state.
In the general case, the ket of any pure optical state, including that of a single photon with extended time-distribution, can be expressed as a linear combination of the kets corresponding to the $\hat{\rho}_k$ defined in Eq.~\eqref{rhok}~\cite{SilberhornMultimode}.

Given the input state $\hat{\rho}_k$, \latin{a priori}, each of the $k$ input photons can generate a detection signal at a time $T_j=t_j+\tau_j$, where the $\tau_j$ are scattered according to the distribution $\vartheta (\tau)$.
However, when a first detection signal is emitted, the detector remains blind over its dead-time~\cite{HadfieldNatPhot2009}.
Consequently, a detector hit by multiple photons delivers only a single ON signal at a time $T=\min(T_j)$ and all detection signals at times longer than T and shorter than the dead-time do not contribute to the detector output.
Note that there is no need to explicitly introduce a dead-time parameter to derive the POVM operators: we consider that after the dead-time, the detector is reset and a new detection cycle can start.
By taking into account the combined effect of dead-time and timing-jitter, $p_{\text{on},k}(T)$ can be expressed as:
\begin{equation}
p_{\text{on}, k}(T) = \sum\limits_{i=1}^k p_{i}(T) \prod\limits_{j \neq i} P\left( t_j + \tau_j > T \right).
\label{eq:p_non_simultane}
\end{equation}
$p_{i}(T)$ is the density probability that photon ``$i$'', corresponding to the minimum of possible signal times (\latin{i.e.\@} $T\equiv T_i$), is actually detected at time $T$.
It can be written as a function of the detection efficiency $\eta$ and of $\vartheta(\tau)$ as $p_{i}(T) = \eta \hspace{0.1cm} \vartheta\left( T-t_i \right)$.
The sum over ``$i$'' states that, \latin{a priori}, every photon can be the first detected.
The product over $j\neq i$ gives the probability that all other photons ``$j$'' are detected at times $T_j > T$, or never detected.
$P\left( t_j + \tau_j > T \right)$ can thus be written as
\begin{equation}
P\left( t_j + \tau_j > T \right) = 1 - \eta \int\limits_{-\infty}^{T} \vartheta(T'-t_j) \diff{T'}.
\label{Pj}
\end{equation}
By putting together Eqs.\,\eqref{eq:pi_diagonal},~\eqref{eq:p_non_simultane} and~\eqref{Pj}, we can derive the general expression for the $\hat{\pi}_{\text{on}}(T)$:
\begin{equation}
\label{eq:pi_non_simultane}
\hat{\pi}_{\text{on}}(T)
= \sum\limits_{k \in \N} \sum\limits_{i = 1}^k \eta \hat{\vartheta}_i(T)\bigotimes\limits_{j \neq i} \left[ \identite - \eta \int\limits_{-\infty}^{T} \hat{\vartheta}_j(T') \diff{T'} \right],
\end{equation}
with the $\hat{\vartheta}_j(T)$ now taking the form of operators:
\begin{equation}
\hat{\vartheta}_{j}(T) = \int \vartheta(T-t_j) \ketbra{1_{t_j}} \diff{t_j}.
\label{eq:def_vartheta_op}
\end{equation}
In the previous expression, we keep different labels for the $\hat{\vartheta}_{j}(T)$ to explicitly recall to which subspace (\latin{i.e.\@} photon) the operator refers to.
However, note that $\hat{\pi}_{\text{on}}(T)$ is independent on the photons' arrival times, $t_j$ being only an integration variable.
We emphasize that, under the assumption of dark-count effects independent on the input state and constant over time, as is the case for most experimental situations, detection noise can be included in the model by simply adding to Eq.~\eqref{eq:pi_non_simultane} the identity operator, $\hat{1}$, weighted by the dark-count rate.

\begin{figure}[htbp]
\centering
\includegraphics[width=.8\columnwidth]{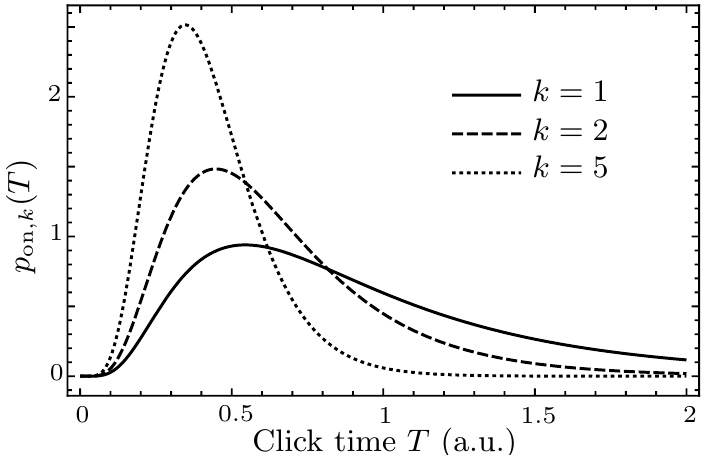}
\caption{\label{fig:p_vs_tau_vs_k}Firing time probability density function for $k=1, 2, 5$ photons arriving simultaneously at time $t=0$ on a lossless detector with $\eta=1$ for a log-normal distribution with mean value $1$ and standard deviation $\frac{1}{2}$.}
\end{figure}

The density operator of Eq.\,\eqref{eq:pi_non_simultane} refers to the general case of non simultaneous input photons.
The special case in which all photons, or a part of them, are simultaneous is obtained by taking $t_j=t$ for $j=1,\ldots,k'$ with $k'\leq k$.
Interestingly, in this case, the jitter effect depends on the number of simultaneous photons $k'$.
Fig.\,\ref{fig:p_vs_tau_vs_k} shows, for different $k$ values, the behavior of $p_{\text{on},k}(T)$ in the limit case of $k'=k$.
As an example, we choose for $\vartheta(\tau)$ a log-normal distribution, with mean value $1$ and standard deviation $\frac{1}{2}$; similar results are obtained for different $\vartheta\left( \tau \right)$ shapes.
Moreover, in order to focus on time effects only, we assume $\eta = 1$.
The case $k = 1$ corresponds to $p_{\text{on},1}(T) = \vartheta\left( T-t \right)$.
As can be seen, for $k \ge 2$, \latin{i.e.\@} for an increasing number of simultaneous photons, the shape of $p_{\text{on},k}(T)$ is modified and shifts towards lower $T$ values.
This confirms an intuitive behavior: the higher the number of photons simultaneously sent to it, the higher the probability for the detector to fire early.
As a consequence, the jitter effect is not the same when the detector is enlightened by few or multiple photons.
We note however that, in standard experimental situations, low $k$ values $\le2$ are generally considered.

\section{Applications}

In the following, we discuss some general applications of the temporal density of POVM operators.
We consider first the direct detection of pure single-photon states of the shape $\ket{\psi} = \int \psi(t)\ket{1_{t}}\diff{t}$~\cite{SilberhornMultimode}, with density matrix $\hat{\rho}=\ketbra{\psi}$, where the width of the temporal distribution $\psi(t)$ is not negligible compared to the amplitude of the detection timing-jitter.
In this condition, temporal uncertainties due to the photon's time delocalization mix with those due to the detection.
This situation corresponds to the case of a photon gun, like for instance an ideal quantum dot operating at low repetition rate~\cite{Senellart}, for which the jitter is much shorter than the time interval between subsequently emitted photons.
From Eq.\,\eqref{eq:pi_non_simultane} for $k=1$, the POVM density expression simply reads
\begin{equation}
\hat{\pi}_{\text{on}}(T)
= \eta \int \vartheta(T-t) \ketbra{1_{t}} \diff{t},
\label{POVM1photon}
\end{equation}
and by applying Eq.\,\eqref{definizione},
\begin{equation}
p_{\text{on}}(T, \ketbra{\psi})
= \eta \int \vartheta(T-t) |\psi(t)|^2 \diff{t}.
\end{equation}
The probability density to detect a photon is thus given by the convolution of the detector temporal response and the probability density of the photon arrival time.

\begin{figure}[htbp]
\centering
\includegraphics[width=.8\columnwidth]{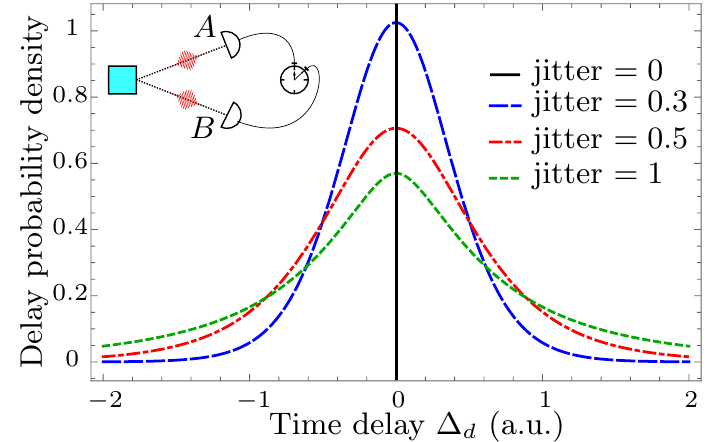}
\caption{\label{fig:coincidence_peak_vs_number_of_pairs}Delay probability density function for a pair of simultaneous photons for different timing-jitters, that refer to different standard deviations of the log-normal distribution.}
\end{figure}

In many situations, experimentalists investigate time correlation measurements between spatially separated photons (see Fig.\,\ref{fig:coincidence_peak_vs_number_of_pairs} inset).
We denote the photons' spatial modes as $A$ and $B$ and indicate their joint state as $\hat{\rho}_{A,B}=\ketbra{\varphi_{A,B}}$, with:
\begin{equation}
\label{eq:etat_paire}
\ket{\varphi_{A,B}} = \int \varphi(t_A, t_B) \ket{t_A}_A \otimes \ket{t_B}_B \diff{t_A}\diff{t_B}.
\end{equation}
The label $\ket{..}_A$ indicated the spatial mode $A$ and analogously for $B$, and $\varphi(t_A, t_B)$ is the so-called joint temporal amplitude~\cite{SilberhornMultimode}.
This is, for instance, the situation at the output of non-degenerate spontaneous parametric down converter (SPDC) where entangled photons can be spatially separated.

Time correlation is evaluated in terms of the joint probability density $p_{\text{on}}(T_A, T_B) = \bra{\varphi_{A,B}} \hat{\pi}_{\text{on}}(T_A, T_B)\ket{\varphi_{A,B}}$, where $\hat{\pi}_{\text{on}}(T_A, T_B) = \hat{\pi}_{A}(T_A) \otimes \hat{\pi}_{B}(T_B)$ is the POVM density extended to the case of two spatially non-degenerated photons.
In order to simplify the result interpretation, we consider only one photon per spatial mode and take for $\hat{\pi}_{A}(T_A)$ and $\hat{\pi}_{B}(T_B)$ the expression of Eq.\,\eqref{POVM1photon}; multiple photon-pair emissions can be easily included by applying, for each mode, the POVM density operator of Eq.\,\eqref{eq:pi_non_simultane}.
We consider two ON/OFF detectors with different timing-jitter functions.
In the limit of one photon per spatial mode,
\begin{multline}
p_{\text{on}}(T_A, T_B) = \eta^2 \iint \vartheta_A(T_A-t_A)\\
\vartheta_B(T_B-t_B) |{\varphi(t_A, t_B)}|^2 \diff{t_A}\diff{t_B}.
\label{eq:p_paires}
\end{multline}
The probability density function expressed by Eq.\,\eqref{eq:p_paires} is a double convolution between the two-detector response functions and the joint temporal intensity of the two-photon state.
It can be experimentally measured by means of time-tagging electronics at the output of the detectors, or by more sophisticated techniques~\cite{TovstonogPRL2008JointTemporalDensity}.
At the same time, in many experimental situations involving pairs of photons, only the measurement of the delay $\Delta_d$ between ``start'' and ``stop'' photons' detection times is required~\cite{JWPanSwapping, Tanzillireview, Lutfi2015}.
The associated probability density is given by:
\begin{equation}
p_{B-A}(\Delta_d)
= \int p_{\text{on}}(T, T + \Delta_d) \diff{T}.
\end{equation}
It can be obtained by application of a POVM density $\hat{\pi}_{B-A}(\Delta_d)$ directly associated with the delay measurement and defined as:
\begin{equation}
\hat{\pi}_{B-A}(\Delta_d)
= \int \hat{\pi}_{A, B}(T, T + \Delta_d) \diff{T}.
\end{equation}
By identifying $T_A= t_A+ \tau$ and $t_B = t_A+ \bar{t}$, $\hat{\pi}_{B-A}(\Delta_d)$ can be conveniently expressed as the convolution of two correlation functions, where the first only depends on the detectors and the second on the input state:
\begin{multline}
\hat{\pi}_{B-A}(\Delta_d)
= \eta^2 \int
\int \vartheta_A(\tau) \vartheta_B(\tau+\Delta_d-\bar{t}) \diff{\tau}\\
\int \ketbra{t_A} \otimes \ketbra{t_A+\bar{t}} \diff{t_A} \diff{\bar{t}}.
\label{POVMdelai}
\end{multline}

For quasi-simultaneous photons, such as for SPDC processes, the photon's joint temporal amplitude takes the form $\varphi(t_A, t_B)= \psi(t_A) \chi(t_B - t_A)$, where $\psi$ is the temporal distribution of the pair emission times and $\chi$ the temporal distribution of the delay in the emission times of two photons for a given pair~\cite{JointTemporalAmplitude}.
Both functions are normalized so that the integral of their square modulus is equal to 1.
By applying Eq.\,\eqref{POVMdelai}, we obtain:
\begin{equation}
p_{B-A}(\Delta_d)
= \eta^2 \int {|\chi(\bar{t})|}^2 \int \vartheta_A(\tau) \vartheta_B(\tau+\Delta_d-\bar{t}) \diff{\tau}\diff{\bar{t}}.
\end{equation}
Here the temporal distribution of the pair emission times disappears, as, in a delay measurement, only the time delay distribution between paired photons is relevant.
Interestingly, in the case of perfectly simultaneous photons, the temporal delay distribution tends to a Dirac's delta distribution ${|\chi(\bar{t})|}^2 = \delta(\bar{t})$ and the time delay for a coincidence event only depends on the detectors' properties.
This can be written as:
\begin{equation}
p_{B-A}(\Delta_d)
= \eta^2 \int \vartheta_A(\tau) \vartheta_B(\tau+\Delta_d) \diff{\tau}.
\end{equation}
As commonly observed in experiments, the shape of the obtained coincidence peaks ($\Delta_d=0$) depends on the timing-jitter~\cite{HadfieldNatPhot2009,ZbindenIJoSTiQE2007SinglePhotonDetection}.
This behavior is shown in Fig.\,\ref{fig:coincidence_peak_vs_number_of_pairs}, where the delay probability density function $p_{B-A}(\Delta_d)$ for a pair of perfectly simultaneous photons is calculated for different jitter standard deviations in the case of two identical detectors with unitary efficiency.

\begin{figure}[htbp]
\centering
\includegraphics[width=.9\columnwidth]{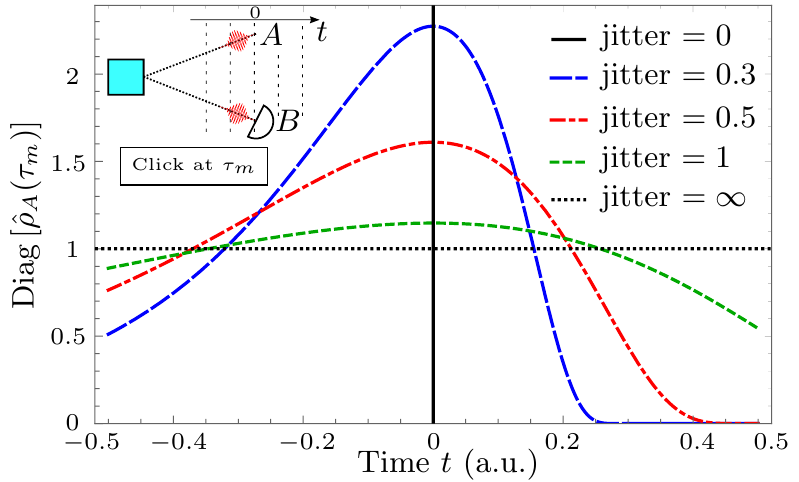}
\caption{\label{fig:Fidelity}Density of diagonal elements of the heralded state $\hat{\rho}_{A}$ as a function of time for different timing-jitter standard deviations. The source emits pairs of simultaneous photons having a rectangular temporal distribution $\psi(t)$ centered in 0 and a width equal to 1. Normalization ensures the area under each curve to be $1$.}
\end{figure}

\smallskip
Eventually, we briefly discuss the effect of the heralding detector timing-jitter in a heralded single photon source based on an entangled state $\hat{\rho}_{A,B}$ produced by SPDC\@.
According to the scheme shown in the inset of Fig.\,\ref{fig:Fidelity}, a photon detection at time $T$ on channel B heralds the presence on channel A of the single photon state described by the density matrix~\cite{DAuriaPRL2011}:
\begin{equation}
\hat{\rho}_A(T) = \frac{\Tr_B\left( \left[ \hat{\identite}_A \otimes \hat{\pi}_B(T) \right] \hat{\rho}_{A,B} \right)}{\Tr\left( \left[ \hat{\identite}_A \otimes \hat{\pi}_B(T) \right] \hat{\rho} _{A,B} \right)},
\end{equation}
where $\Tr_B$ is the partial trace with respect to subsystem B.
From Eqs.\,\eqref{eq:etat_paire} and~\eqref{POVM1photon}, in case of simultaneous photons (${|\chi(\bar{t})|}^2 = \delta(\bar{t})$), we obtain:
\begin{equation}
\label{eq:rho_annonce}
\hat{\rho}_A(T) = \frac{\int
	\vartheta(T-t)
	{|\psi(t)|}^2
	\ketbra{1_{t}}
	\diff{t}}{\int	\vartheta(T-t) {|\psi(t)|}^2 \diff{t}},
\end{equation}
where, as before, $\psi$ refers to the temporal distribution of the pair.
As shown in Eq.\,\eqref{eq:rho_annonce}, the heralded state is a mixture represented by a diagonal density operator: the ensemble of diagonal elements represents the density probability function for the heralded photon emission time.
Fig.\,\ref{fig:Fidelity} represents them as functions of time $t$, for different timing-jitters and for perfectly simultaneous photons in channels A and B.
We assume that the heralding detector fires at time $T=0+\tau$.
In the ideal case of a detector with no jitter, $\vartheta(\tau)\rightarrow\delta(\tau-\tau_m)$, with $\tau_m$ a constant detection delay.
As expected, the heralded state is perfectly defined and found in the pure state $\ketbra{1_{0}}$: its density matrix has only one non-zero element.
Conversely, for infinite jitter (\latin{i.e.\@} for an infinitely large $\vartheta(\tau)$), no information is retrieved from the heralding time and the heralded state is a mixture with a flat density probability function.
We note that in the case of multiple pair emissions, where more than two simultaneous photons are present on channels A and B, the purity of the state is further decreased by the inability of ON/OFF detectors to resolve the number of photons~\cite{DAuriaPRL2011}.

\section{Conclusion}
We have presented an operational model, exploiting the POVM formalism to describe the characteristics of an ON/OFF single photon detector with non negligible timing-jitter and in presence of dead-time.
Although we mostly provide simple examples involving one or two photons impinging on the detectors, our formalism is capable of describing general experimental situations.
We therefore believe that these characteristics stand as a valuable help for a better comprehension of the timing-jitter effect in the perspective of developing ultra-fast quantum communication.

\begin{acknowledgments}
The authors acknowledge financial support from the Agence Nationale de la Recherche (SPOCQ ANR-14-CE32-0019-03, CONNEQT ANR-2011-EMMA-0002, HyLight ANR-17-CE30-0006-01).
A.\,Z.\ acknowledges support from the CNR Short-Term Mobility Program and the Université Nice Sophia Antipolis for an invited professor fellowship.

É.G. and B.F. contributed equally to this work.
\end{acknowledgments}

%

\end{document}